\documentclass[prb,aps,twocolumn,preprintnumbers,amsmath,amssymb,superscriptaddress,showpacs]{revtex4-2}

\usepackage{graphicx,xfrac}
\usepackage{amsmath}
\usepackage{amssymb}
\usepackage{multirow}
\usepackage{ulem}
\usepackage{float}
\usepackage[colorlinks=true,linkcolor = blue,citecolor=blue,urlcolor=blue,anchorcolor = blue]{hyperref}
\usepackage[sort&compress]{natbib}
\usepackage{times}
\usepackage{bm}
\usepackage[version=4]{mhchem}

\usepackage{tabularx}
\usepackage{colortbl}
\usepackage[table,xcdraw]{xcolor}
\usepackage{multirow}
\usepackage{array}
\newcommand{\rows}[1]{\multirow{2}*{#1}}
\newcommand{\cclrA}{\cellcolor[HTML]{B4DD2B}}
\newcommand{\cclrB}{\cellcolor[HTML]{DBEE97}}
\newcommand{\cclrC}{\cellcolor[HTML]{FDE724}}
\newcommand{\cclrD}{\cellcolor[HTML]{FEF38F}}
\newcommand{\cclrE}{\cellcolor[HTML]{FEF8BC}}
\newcommand{\cclrF}{\cellcolor[HTML]{808080}}
\newcommand{\cclrG}{\cellcolor[HTML]{A6A6A6}}
\newcommand{\cclrH}{\cellcolor[HTML]{BFBFBF}}
\newcommand{\cclrI}{\cellcolor[HTML]{D9D9D9}}

\begin{document}
\title{Rare-Earth Chalcogenides: An Inspiring Playground For Exploring Frustrated Magnetism}

\author{Mingtai\,Xie}
\affiliation{School of Physical Science and Technology, Lanzhou University, Lanzhou 730000, China}
\affiliation{Beijing National Laboratory for Condensed Matter Physics, Institute of Physics, Chinese Academy of Sciences, Beijing 100190, China}

\author{Weizhen\,Zhuo}
\affiliation{School of Physical Science and Technology, Lanzhou University, Lanzhou 730000, China}
\affiliation{Beijing National Laboratory for Condensed Matter Physics, Institute of Physics, Chinese Academy of Sciences, Beijing 100190, China}

\author{Yanzhen\,Cai}
\affiliation{School of Physical Science and Technology, Lanzhou University, Lanzhou 730000, China}
\affiliation{Beijing National Laboratory for Condensed Matter Physics, Institute of Physics, Chinese Academy of Sciences, Beijing 100190, China}

\author{Zheng\,Zhang}
\email[e-mail:]{zhangzheng@iphy.ac.cn}
\affiliation{Beijing National Laboratory for Condensed Matter Physics, Institute of Physics, Chinese Academy of Sciences, Beijing 100190, China}

\author{Qingming\,Zhang}
\email[e-mail:]{qmzhang@iphy.ac.cn}
\affiliation{School of Physical Science and Technology, Lanzhou University, Lanzhou 730000, China}
\affiliation{Beijing National Laboratory for Condensed Matter Physics, Institute of Physics, Chinese Academy of Sciences, Beijing 100190, China}

\begin{abstract}
	
	The rare-earth chalcogenide $ARECh_{2}$ family ($A$\,=\,alkali metal or monovalent ions, $RE$\,=\,rare earth, $Ch$\,=\,chalcogen) has emerged as a paradigmatic platform for studying frustrated magnetism on a triangular lattice. The family members exhibit a variety of ground states, from quantum spin liquid to exotic ordered phases, providing fascinating insight into quantum magnetism. Their simple crystal structure and chemical tunability enable systematic exploration of competing interactions in quantum magnets. Recent neutron scattering and thermodynamic studies have revealed rich phase diagrams and unusual excitations, refining theoretical models of frustrated systems. This review provides a succinct introduction to $ARECh_{2}$ research. It summarizes key findings on crystal structures, single-ion physics, magnetic Hamiltonians, ground states, and low-energy excitations. By highlighting current developments and open questions, we aim to catalyze further exploration and deeper physical understanding on this frontier of quantum magnetism.

\end{abstract}

\maketitle


\section{Introduction}
\label{section:Intro}
	Quantum spin liquid (QSL) is a key concept in modern condensed matter physics, first proposed by Anderson in 1973 \cite{ANDERSON1973153}. It describes a highly entangled quantum state of a spin system that lacks long-range magnetic order even at zero temperature, driven by strong quantum fluctuations. Anderson initially introduced this concept while considering the antiferromagnetic Heisenberg model on a triangular-lattice antiferromagnet (TLAF). In this idealized model, due to geometric frustration, the spins cannot satisfy all interactions simultaneously, potentially resulting in a highly entangled ground state with long-range quantum entanglement \cite{balents_spin_2010}. This state does not exhibit conventional magnetic order but rather displays liquid-like behavior with ubiquitous and dynamic spin fluctuations \cite{doi:10.1126/science.1163196}. The characteristics of QSLs make them an ideal platform for understanding highly entangled quantum systems and have potential significance for exploring quantum information technology and understanding phenomena like high-temperature superconductivity \cite{RevModPhys.89.025003}.
	
	Since Anderson's proposal of the QSL, extensive studies have been conducted on the Heisenberg model on TLAFs to explore the possibility of realizing such a state \cite{doi:10.1080/14786439808206568}. Historically, the Heisenberg model on a TLAF is known for its 120° ordered ground state, a characteristic arrangement where each spin forms a 120° angle with its neighbors \cite{lee1984DiscreteSymmetry, kawamura1984Phase}. This geometrically frustrated configuration was initially regarded as a potential candidate for observing a QSL due to the inherent inability of the spins to simultaneously satisfy all pairwise interactions.
	Several alternative spin models on TLAFs have been proposed and examined to achieve the elusive QSL. These models include the XXZ model \cite{PhysRevLett.112.127203} and the $J_{1}-J_{2}$ model \cite{PhysRevB.92.041105}. Despite the introduction of these additional complexities, the quest to definitively achieve a QSL state in these systems remains a challenge and points towards the need for either novel theoretical approaches or the discovery of new material systems with unique interaction schemes.
	
	With the progress of these theoretical models, research in TLAF materials primarily concentrates on the transition metal compounds. For example, the organic TLAF materials \ce{\kappa-(BEDT-TTF)2Cu2(CN)3} \cite{doi:10.1126/science.abc6363,Yamashita2009,Isono2016} and \ce{EtMe3Sb[Pd(dmit)2]2} \cite{PhysRevLett.123.247204,PhysRevB.84.094405,PhysRevB.101.140407} are considered promising candidates for hosting QSLs. In particular, thermal conductivity experiments on \ce{\kappa-(BEDT-TTF)2Cu2(CN)3} reveal a non-zero thermal conductivity as the temperature approaches 0\,K, suggesting the presence of itinerant spinon excitations.
	\ce{Na2BaCo(PO4)2} \cite{doi:10.1073/pnas.1906483116} was also initially considered as a candidate material for QSL, but more detailed experimental measurements and theoretical studies have shown that its ground state is a spin supersolid \cite{Gao2022, Xiang2024}.
	On the other hand, transition metal TLAF materials present experimental challenges due to their complex compositions, inhomogeneities, and difficulties in obtaining high-quality single crystals. These factors have complicated efforts to definitively identify QSL states in these materials.
	
	Triangular-lattice rare-earth magnets have recently drawn significant interest due to their inherent geometric frustration and strong spin-orbit coupling (SOC), which facilitate anisotropic spin-exchange interactions as described by the Jackeli--Khaliullin scenario  \cite{PhysRevLett.101.216804,PhysRevLett.105.027204}. Particularly, rare-earth ions with an odd number of 4$f$ electrons form stable Kramers doublets, protected by time reversal symmetry and resilient to lattice distortions and structural defects.
	In 2015, we discovered \ce{YbMgGaO4} \cite{Li2015,PhysRevLett.118.107202}, a rare-earth TLAF that meets the aforementioned criteria. A large amount of experimental evidence, including inelastic neutron scattering (INS) \cite{Paddison2017,PhysRevLett.122.137201}, muon spin relaxation ($\mu$SR) \cite{PhysRevLett.117.097201}, and thermodynamic measurements, consistently supports that the ground state of \ce{YbMgGaO4} is a gapless QSL state.
	Therefore, \ce{YbMgGaO4} is considered as a typical QSL material.

	Following this discovery, in 2018 we introduced the rare-earth chalcogenides $ARECh_2$ ($A$\,=\,alkali or monovalent ions, $RE$\,=\,rare earth, $Ch$\,=\,chalcogen) \cite{liu2018RareEarth} as a new and extensive family of candidates for triangular-lattice QSLs.
	These materials, with their simple structure and chemical formula, remove the issue on possible exchange disorders and have become the most representative examples of TLAFs. Exploration of the $ARECh_2$ family has yielded intriguing results. 
	For example, Yb-based compounds such as \ce{NaYbO2} \cite{ding2019Gapless, bordelon2019Fieldtunablea, PhysRevB.99.180401, PhysRevB.108.L140411}, \ce{NaYbS2} \cite{baenitz2018NaYbS2,wu2022Magnetic,zhuo2024Magnetism,sarkar2019Quantum,sichelschmidt2019Electron,haussler2022Diluting}, and \ce{NaYbSe2} \cite{ranjith2019Anisotropic,zhang2022Lowenergya,dai2021Spinon,scheie2024Spectrum,scheie2024Nonlinear,zhang2021Crystalline,zhang2021Effective} were proposed to host QSL states.
	The ground states of \ce{KYbSe2} \cite{scheie2023Proximate} and \ce{CsYbSe2} \cite{xie2023Complete} have been confirmed to be 120° ordered states.
	In \ce{Er}-based and \ce{Ce}-based compounds, such as \ce{KErSe2} \cite{xing2021Stripe}, \ce{KCeS2} \cite{avdoshenko2022Spinwave}, and \ce{CsCeSe2} \cite{xie2023Stripe}, the ground states are characterized by stripe-$x$ or stripe-$yz$ ordering.
	Tm-based compounds, such as \ce{KTmSe2} \cite{zheng2023Exchangerenormalized}, exhibit Ising-type spin characteristics, which can be described by the transverse-field Ising model (TFIM). 
	The variety of ground states in this family provide a robust platform for exploring the complexities of frustrated magnetism.
	
	In this work, we provide a brief review of recent research on rare-earth chalcogenides $ARECh_2$, outlining key paradigms to help newcomers quickly integrate into this field. The review is structured into the following parts:
	In Section \ref{section:Struc}, we begin by examining the diverse crystal structures within this family and discuss methodologies for crystal synthesis, establishing the foundational aspects of these materials.
	Moving to Section \ref{section:Ion}, the focus shifts to the single-ion magnetism of rare-earth ions and the significant role of the crystal electric field (CEF), bridging structural characteristics with magnetic properties.
	In Section \ref{section:Finite}, we introduce the Hamiltonian models typically used to describe the interactions within this system, linking the microscopic quantum magnetic mechanisms to the results of experimental measurements.
	Section \ref{section:GS} delves into the typical ground states exhibited by family members, exploring the consequences of these interactions and the resulting magnetic phenomena.
	Finally, in Section \ref{section:Outlook}, we consider future research directions, aiming to inspire further developments and uncover new insights within this intriguing field of study.

\begin{figure*}[ht]
	\includegraphics[width=1\linewidth]{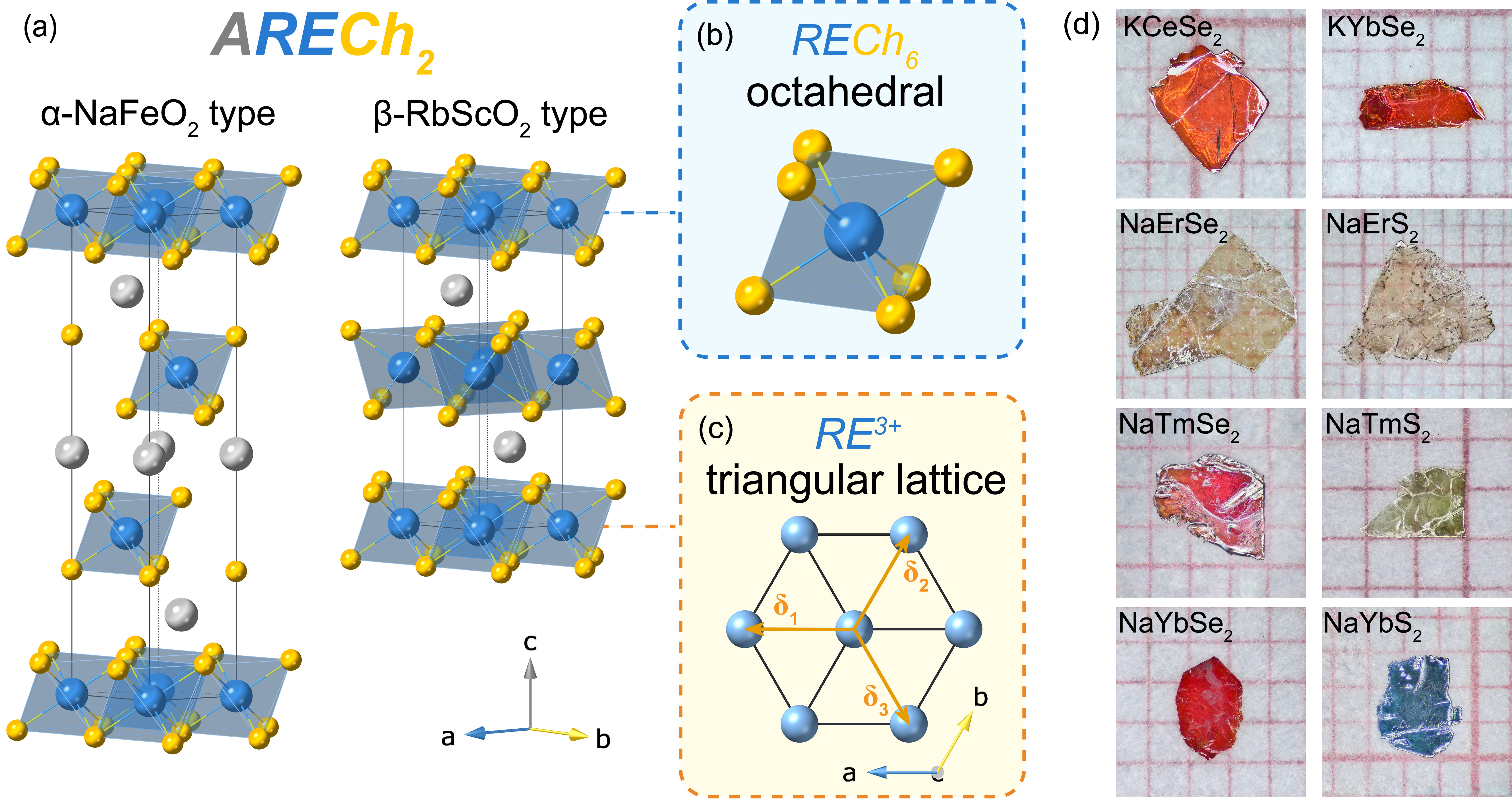}
	\caption{\label{fig:structure}  \textbf{Typical crystal structures and single crystals of $\bm {ARECh_{2}}$ ($\bm{A}$\,=\,alkali, $\bm{RE}$\,=\,rare earth, $\bm{Ch}$\,=\,chalcogen).} (a) The $\alpha$-NaFeO$_{2}$-type structure (space group $R\bar{3}m$) and the $\beta$-RbScO$_{2}$-type structure (space group $P6_{3}/mmc$), both share the same layered structure, featuring (b) $RECh_{6}$ octahedra arranged in (c) a triangular lattice. The $\bm{\delta}_{\alpha}\  (\alpha=1,2,3)$ vectors represent the three primitive vectors for bond-dependent interactions. (d) Photographs of representative $ARECh_{2}$ single crystals. The grid is in 1 mm scale.
	}
\end{figure*}

\section{Structure}
\label{section:Struc}
\begin{table}[h]
	\caption{Structure map of the $ARECh_{2}$ ($A$\,=\,alkali metal, $RE$\,=\,rare earth, $Ch$\,=\,chalcogen) family}
	\label{table:struc}
	\raggedright 
	\resizebox{0.9\columnwidth}{!}{%
		\parbox{\columnwidth}{
			\setlength{\tabcolsep}{0pt}
			\renewcommand{\arraystretch}{0.7}
			\newcolumntype{M}{>{\centering\arraybackslash}m{0.5cm}}
			\newcolumntype{N}{>{\centering\arraybackslash}m{0.4cm}}
			\newcolumntype{O}{>{\centering\arraybackslash}m{1.5cm}}
			\begin{tabular}{|M|M|M|M|M|M|M|M|M|M|M|M|M|M|M|M|O|}
				\hline
				\multicolumn{17}{|c|}{$ARE$O$_{2}$}\rule{0pt}{10pt}\\
				\hline
				\rows{} & \rows{La} & \rows{Ce} & \rows{Pr} & \rows{Nd} & \rows{Sm} & \rows{Eu} & \rows{Gd} & \rows{Tb} & \rows{Dy} & \rows{Ho} & \rows{Er} & \rows{Tm} & \rows{Yb} & \rows{Lu} & \rows{Y} & \rows{Ref.} \\ &&&&&&&&&&&&&&&&\\ 
				\hline
				Li 
				& \cclrI & \cclrI & \cclrI & \cclrI & \begin{tabular}{@{}M@{}} \cclrI \\ \cclrG \end{tabular} & \cclrG & \cclrG & \begin{tabular}{@{}M@{}} \cclrI \\ \cclrG \end{tabular} & \cclrI & \cclrI & \cclrF & \cclrF & \cclrF & \cclrF & \cclrI 
				&\cite{liu2018RareEarth, cantwell2011Crystal, hashimoto2002Structures} \\ \hline
				Na 
				& \cclrF & \cclrF & \cclrF & \cclrF & \cclrF & \cclrF & \cclrF &  & \cclrH & \cclrH & \begin{tabular}{@{}M@{}} \cclrH\\ \cclrB \end{tabular} & \cclrB & \cclrB & \cclrB & \cclrB 
				&\cite{liu2018RareEarth, blasse1966Sodium, hashimoto2003Magnetic} \\ \hline
				K 
				& \begin{tabular}{@{}M@{}} \cclrB \\ \cclrB \end{tabular} & \cclrB & \cclrB & \cclrB & \cclrB & \cclrB & \cclrB & \cclrB & \cclrB & \cclrB & \cclrB & \cclrB & \cclrB & \cclrB & \cclrB 
				& \cite{liu2018RareEarth, dong2008Structure} \\ \hline
				Rb
				& \begin{tabular}{@{}M@{}} \cclrB \\ \cclrB \end{tabular} & \cclrB &  & \cclrB & \cclrB & \cclrB & \cclrB &  & \cclrB & \cclrB & \cclrB & \cclrB & \cclrB & \cclrB &
				& \cite{seeger1969Ternaere, brunn1975Neue, ortiz2022Electronic}
				\\ \hline
				Cs
				&  &  &  & \begin{tabular}{@{}M@{}} \cclrA \\ \cclrA \end{tabular} &  &  &  &  &  &  &  &  &  &  &  
				& \cite{xing2024Candidate} \\  
				\hline
				
				\multicolumn{17}{|c|}{$ARE$S$_{2}$}\rule{0pt}{10pt}\\ 
				\hline
				Li 
				& \cclrE & \cclrE & \cclrD & \cclrD & \cclrD & \cclrD & \cclrD & \cclrD & \begin{tabular}{@{}M@{}} \cclrD\\ \cclrB \end{tabular} & \begin{tabular}{@{}M@{}} \cclrD\\ \cclrB \end{tabular} & \begin{tabular}{@{}M@{}} \cclrD\\ \cclrB \end{tabular} & \cclrB & \cclrB & \cclrB
				& \begin{tabular}{@{}M@{}} \cclrD\\ \cclrB \end{tabular}
				& \cite{ohtani1987Synthesis, fabry2014Structure, ballestracci1965Etude} \\ 
				\hline
				Na 
				& \cclrD & \cclrD & \cclrD & \begin{tabular}{@{}M@{}} \cclrD\\ \cclrB \end{tabular} & \begin{tabular}{@{}M@{}} \cclrD\\ \cclrB \end{tabular} & \cclrB & \cclrB & \cclrB & \cclrB & \cclrB & \cclrB & \cclrB & \cclrB & \cclrB & \cclrB
				& \cite{fabry2014Structure, ballestracci1964Etude, sato1984Preparation, verheijen1975FLUX, schleid1993ChemInform} \\ 
				\hline
				K 
				& \begin{tabular}{@{}M@{}} \cclrB \\ \cclrB \end{tabular} & \cclrB & \cclrB & \cclrB & \cclrB & \cclrB & \cclrB & \cclrB & \cclrB & \cclrB & \cclrB & \cclrB & \cclrB & \cclrB & \cclrB 
				& \cite{fabry2014Structure, ballestracci1965Etude} \\ \hline
				Rb 
				& \begin{tabular}{@{}M@{}} \cclrB \\ \cclrB \end{tabular} & \cclrB & \cclrB & \cclrB & \cclrB & \cclrB & \cclrB & \cclrB & \cclrB & \cclrB & \cclrB & \cclrB & \cclrB & \cclrB & \cclrB 
				& \cite{fabry2014Structure, bronger1996ChemInform} \\ \hline
				Cs 
				& \cclrB & \cclrB & \begin{tabular}{@{}M@{}}\cclrB \\ \cclrA \end{tabular} & \begin{tabular}{@{}M@{}}\cclrB \\ \cclrA \end{tabular} & \begin{tabular}{@{}M@{}}\cclrB \\ \cclrA \end{tabular} & \begin{tabular}{@{}M@{}}\cclrB \\ \cclrA \end{tabular} & \begin{tabular}{@{}M@{}}\cclrB \\ \cclrA \end{tabular} & \begin{tabular}{@{}M@{}}\cclrB \\ \cclrA \end{tabular} & \begin{tabular}{@{}M@{}}\cclrB \\ \cclrA \end{tabular} & \begin{tabular}{@{}M@{}}\cclrB \\ \cclrA \end{tabular} & \begin{tabular}{@{}M@{}}\cclrB \\ \cclrA \end{tabular} & \begin{tabular}{@{}M@{}}\cclrB \\ \cclrA \end{tabular} & \begin{tabular}{@{}M@{}}\cclrB \\ \cclrA \end{tabular} & \begin{tabular}{@{}M@{}}\cclrB \\ \cclrA \end{tabular} & 
				& \cite{bronger1993Zur} \\ 
				\hline
				
				\multicolumn{17}{|c|}{$ARE$Se$_{2}$}\rule{0pt}{10pt}\\ 
				\hline
				Li 
				& \cclrE & \cclrE &  & \cclrD & \cclrD &  & \begin{tabular}{@{}M@{}} \cclrD\\ \cclrB \end{tabular} & \begin{tabular}{@{}M@{}} \cclrD\\ \cclrB \end{tabular} & \begin{tabular}{@{}M@{}} \cclrD\\ \cclrB \end{tabular} & \begin{tabular}{@{}M@{}} \cclrD\\ \cclrB \end{tabular} & \begin{tabular}{@{}M@{}} \cclrD\\ \cclrB \end{tabular} &  & \cclrC &  & 
				\begin{tabular}{@{}M@{}} \cclrD\\ \cclrB \end{tabular}
				& \cite{ohtani1987Synthesis, dissanayakamudiyanselage2022LiYbSe2} \\ 
				\hline
				Na & \cclrD & \begin{tabular}{@{}M@{}} \cclrD\\ \cclrB \end{tabular} &  & \cclrB & \cclrB &  & \cclrB & \cclrB & \cclrB & \cclrB & \cclrB &  & \cclrB & \cclrB & \cclrB
				& \cite{liu2018RareEarth, ohtani1987Synthesis, xing2019Synthesisa} \\ 
				\hline
				K
				& \begin{tabular}{@{}M@{}} \cclrB\\ \cclrB \end{tabular} & \cclrB & \cclrB & \cclrB & \cclrB &  &  &  &  &  &  & \cclrB & \cclrB & \cclrB & \cclrB
				& \cite{xing2019Synthesisa, gray2003Crystal, xing2021Synthesis, sanjeewa2022Synthesis, zheng2023Exchangerenormalized} \\ 
				\hline
				Rb
				& \begin{tabular}{@{}M@{}} \cclrB\\ \cclrB \end{tabular} & \cclrB & \cclrB & \cclrB & \cclrB &  & \cclrB & \cclrB &  & \cclrB & \cclrB &  & \cclrB & \cclrB &   
				& \cite{deng2002New, xing2021Synthesis} \\ 
				\hline
				Cs
				& \begin{tabular}{@{}M@{}} \cclrB\\ \cclrB \end{tabular} & \cclrB & \cclrB & \cclrB & \cclrB &  &  &\cclrA & \cclrA &  & \cclrA & \cclrA & \cclrA & \cclrA &  
				& \cite{deng2005CsYbSe2, xing2020Crystal} \\ 
				\hline
				
				\multicolumn{17}{|c|}{$ARE$Te$_{2}$}\rule{0pt}{10pt}\\ 
				\hline
				\rows{Li} 
				&  &  &  &  &  &  &  &  &  &  &  &  &  &  &  & \\ 
				&  &  &  &  &  &  &  &  &  &  &  &  &  &  &  & \\ 
				\hline
				Na 
				&  &  & \begin{tabular}{@{}M@{}} \cclrC\\ \cclrB \end{tabular} & \cclrB & \cclrB &  &  & \cclrB &  &  &  & \cclrB &  & \cclrB & 
				& \cite{bronger1993Zur, lissner2003Single, eto2023Structural, zheng2024Interplay} \\ 
				\hline
				K
				& \begin{tabular}{@{}M@{}} \cclrB\\ \cclrB \end{tabular} &  & \cclrB & \cclrB & \cclrB &  & \cclrB &  &  &  & \cclrB &  &  &  & \cclrB
				& \cite{bronger1993Zur, stoewe2003Synthesen, keane1992Structure, babo2009Two} \\ 
				\hline
				Rb
				&  & \begin{tabular}{@{}M@{}} \cclrB\\ \cclrB \end{tabular} &  & \cclrB & \cclrB &  &  &  &  &  &  &  &  &  & \cclrA
				& \cite{bronger1993Zur, stoewe2003Synthesen, babo2009Two} \\ 
				\hline
				Cs
				&  &  &  & \begin{tabular}{@{}M@{}} \cclrB\\ \cclrB \end{tabular} &  &  &  &  &  &  &  &  &  &  & 
				& \cite{stoewe2003Synthesen} \\ 
				\hline
			\end{tabular}
			
			\vspace{0.5em}
			
			\footnotesize
			\begin{tabular}{NlNlNl}
				\cclrA & \ P6$_{3}$/mmc, $\beta$-RbScO$_{2}$ type \hspace{0.5cm} &  \cclrC & \ Fm$\bar{3}$m \hspace{0.5cm} & \cclrG & \ Pnma \rule{0pt}{10pt}
				\\ 
				\cclrB & \ R$\bar{3}$m, $\alpha$-NaFeO$_{2}$ type &\cclrE & \ I$\bar{4}$3d \hspace{0.5cm} & \cclrH& \ C2/c \rule{0pt}{10pt}
				\\ 
				\cclrC & \ Fd$\bar{3}$m  & \cclrF & \ I4$_{1}$/amd \hspace{0.5cm} & \cclrI & \ P2$_{1}$/c  \rule{0pt}{10pt}
			\end{tabular}
		}
	}
\end{table}
	
	 In alkali rare-earth chalcogenides $ARECh_2$ ($A$\,=\,alkali metal, $RE$\,=\,rare earth, $Ch$\,=\,chalcogen), there are at least nine crystal structures with different space group symmetries, as listed in Table \ref{table:struc}.
	 The $\alpha$-\ce{NaFeO2}-type structure (space group: R$\bar{3}$m) predominates across the family, while the $\beta$-\ce{RbScO2}-type structure (space group: P6$_{3}$/mmc) is exclusively observed in Cs-based compounds.
	 These two primary structures differ from each other markedly in their layer stacking sequences:
	 the $\alpha$-\ce{NaFeO2}-type features ABCABC stacking, whereas the $\beta$-\ce{RbScO2}-type adopts ABAB stacking, as shown in Fig. \ref{fig:structure}(a).
	 The remaining crystal structures are less common in $ARECh_2$ and are primarily found in those containing Li and Na.

	 Research has predominantly focused on $\alpha$-\ce{NaFeO2}-type and $\beta$-\ce{RbScO2}-type $ARECh_2$ compounds , which share two critical features: 
	 (i) magnetic ions $RE^{3+}$ form perfect two-dimensional triangular lattices [Fig. \ref{fig:structure}(c)] and (ii) the local $RECh_{6}$ octahedra possess inversion symmetry, precluding Dzyaloshinskii--Moriya (DM) interactions [Fig. \ref{fig:structure}(b)]. 
	 These characteristics render $ARECh_2$ compounds ideal systems for studying TLAFs.
	
	 Additionally, there are several important details to note in the rare-earth chalcogenides $ARECh_2$:
	 (i) The $\alpha$-\ce{NaFeO2}-type structure is often confused with the delafossite structure which is present in O-based compounds, with the A site being confined to Cu$^{+}$, Ag$^{+}$, Pd$^{+}$, and Pt$^{+}$  \cite{marquardt2006Crystal}. Distinct from the $\alpha$-NaFeO$_{2}$-type structure, where the $A^{+}$ ion is hexacoordinated, the delafossite structure features $A^{+}$ ions in linear coordination with two oxygen ions. Depending on stacking order, it can also form 3R- and 2H-type structures. In the alkali rare-earth chalcogenides $ARECh_2$, no delafossite structure has been found yet. Notably, Cs$RE$S$_{2}$ ($RE$\,=\, Pr--Lu) can adopt both the 3R- and 2H-type structures \cite{bronger1993Zur}, offering an opportunity to investigate the influence of stacking order on magnetic behavior.
	 (ii) Regarding the role of each component in $ARECh_{2}$, the magnetic properties are primarily governed by $RE^{3+}$ ions arranged in quasi-two-dimensional triangular lattices, with non-magnetic $A^{+}$ ions acting as spacers between these layers.  
	 However, recent experimental results indicate that non-magnetic $A^{+}$ ions also play a subtle yet significant role in modulating the system's magnetism \cite{grussler2023Roleb, xing2021Synthesis, ortiz2022Electronic}.
	 Substitution of $A^{+}$ ions alters the interlayer spacing, which in turn affects the $RECh_{6}$ octahedral geometry. These structural modifications change both the $RE^{3+}$--$Ch^{2-}$--$RE^{3+}$ bond angles and the nearest-neighbor (NN) $RE^{3+}$--$RE^{3+}$ distances, thereby tuning the spin exchange interactions. 
		
	 As for sample synthesis, the majority of the compounds listed in Table \ref{table:struc} can be synthesized into high-quality single crystals.
	 However, an exception exists: O-based compounds have only been synthesized as polycrystalline samples to date.
	 This limitation has hindered in-depth studies of O-based compounds.
	 
	 The synthesis methods for $ARECh_2$ compounds depend heavily on specific chalcogen elements. 
	 For the O-series, powder samples are predominantly synthesized via solid-state reaction using $A_{2}$CO$_{3}$ and $RE_{2}$O$_{3}$ \cite{liu2018RareEarth,hashimoto2003Magnetic,dong2008Structure}. 
	 The S-, Se-, and Te-series powders can be synthesized through solid-state reactions using either $A_{2}Ch$ precursors or elemental mixtures \cite{liu2018RareEarth, xing2019Synthesisa, eto2023Structural}.
	 For single crystals, the S-series can be prepared using either vapor transport \cite{avdoshenko2022Spinwave,bastien2020Longrange,kulbakov2021Stripeyz} or a salt-flux method \cite{fabry2014Structure,sato1984Preparation, baenitz2018NaYbS2}; Se-based compounds are mostly prepared using a salt-flux method \cite{liu2018RareEarth, xing2019Synthesisa, xing2021Synthesis, sanjeewa2022Synthesis, xing2020Crystal}; while Te-based compounds are typically prepared using a Te-flux method \cite{Liu_2021, zheng2024Interplay}. 
	 Some of the single crystal samples synthesized using the methods mentioned above are shown in Fig. \ref{fig:structure}(d). These single crystals exhibit a distinctive flake-like morphology with easily cleavable planes, reflecting their layered structure. Their colorful transparency is indicative of their insulating nature, characterized by electronic band gaps of several electron volts \cite{liu2018RareEarth,Liu_2021}.
	 
	 The synthesis of $ARECh_{2}$ compounds requires standard solid-state reaction equipment: quartz tubes, crucibles, and muffle or tube furnaces. Some compounds necessitate specific gas atmospheres and associated gas-handling systems. This relatively simple experimental setup, which is achievable in most materials laboratories, has facilitated the rapid exploration of the $ARECh_{2}$ family. The accessibility of these synthesis methods has been crucial in accelerating research progress and enabling a broad, systematic study of this diverse material system.

\section{Single-ion magnetism and CEF}
\label{section:Ion}
	
	\begin{figure*}[t] 	
		\includegraphics[width=1\linewidth]{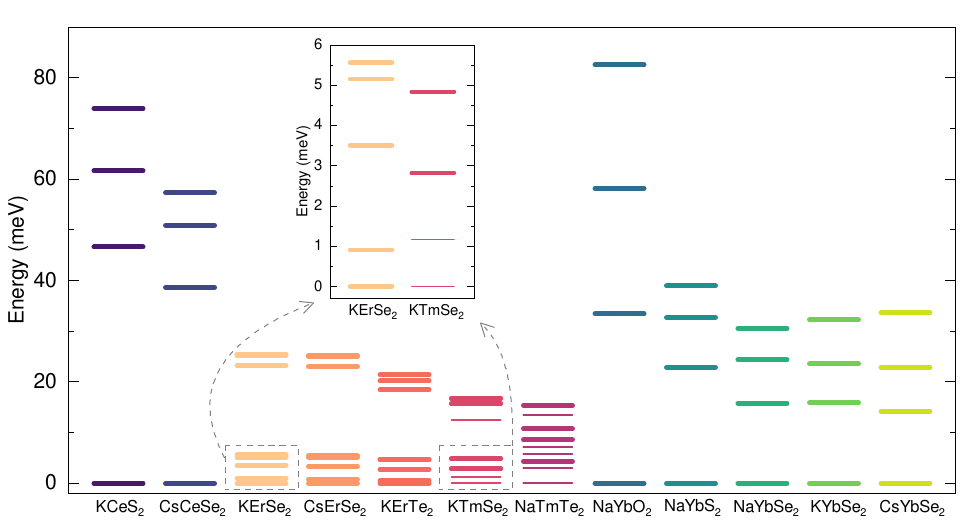}
		\caption{\label{fig:CEF}  \textbf{Schematic diagram of CEF energy levels of $\bm {ARECh_{2}}$.} $RE$\,=\,Ce \cite{bastien2020Longrange, xie2023Stripe}, Er \cite{scheie2020Crystalfield, liu2024Finite}, Tm \cite{zheng2023Exchangerenormalized, zheng2024Interplay}, Yb \cite{bordelon2020Spin, baenitz2018NaYbS2, zhang2021Crystalline, scheie2023Proximate, pai2022Mesoscale}. The inset is a magnified view of KErSe$_{2}$ and KTmSe$_{2}$. All energy levels for Kramers ions are doubly degenerate. In the inset, the thicker solid lines represent doubly degenerate energy levels, and the thinner solid lines represent single energy levels.
		}
	\end{figure*}
	
	In the $ARECh_{2}$ family, rare-earth ions are primarily present in the trivalent state.
	Trivalent rare-earth ions ($RE^{3+}$), with their unique electronic configurations and strong SOC, exhibit fascinating quantum behaviors.
	$RE^{3+}$ typically have the electron configuration of [Xe] 4$f^n$, where $n$ varies from 0 to 14 across the lanthanide series. This configuration results from the filling of the 4$f$ orbital. 
	Due to shielding by the outer 5$s$ and 5$p$ orbitals, the 4$f$ electron orbitals exhibit significant localization.
	Additionally, the substantial mass of the rare-earth ions enhances SOC. These characteristics ensure that the single-ion ground state of $RE^{3+}$ is well described by Hund's rules.
	Furthermore, according to Kramers' theorem, $RE^{3+}$ ions can be categorized into two groups: Kramers ions ($RE$\,=\,Ce, Nd, Sm, Gd, Dy, Er, Yb), and non-Kramers ions ($RE$\,=\,Pr, Eu, Tb, Ho, Tm).
	For Kramers ions, regardless of how the CEF environment of the surrounding anions changes, the energy levels after CEF splitting are protected by time-reversal symmetry and are at least doubly degenerate.
	
	Building on the foundation of single-ion magnetism, we further consider the impact of the surrounding anions, i.e., the CEF, on the magnetism of the $RE^{3+}$ ions.
	As illustrated in Fig. \ref{fig:structure}(b), the $RE^{3+}$ ion is surrounded by $Ch^{2-}$ ions, with the six nearest neighbor $Ch^{2-}$ and $RE^{3+}$ ions forming a distorted $RECh_{6}$ octahedron with $D_{3d}$ symmetry. 
	After SOC, the energy level of a single $RE^{3+}$ ion is further split into multiple energy levels by the CEF.
	Theoretically, CEF theory treats the central valence electron as a quantum system and the surrounding ligand ions as a classical charge system. For a single $RE^{3+}$ ion at an arbitrary lattice site $i$, using the Stevens operators we can write the CEF Hamiltonian as \cite{stevens1952Matrix}:
	\begin{align}
		{\mathcal H_{\rm CEF}}(i) =\ \sum_{m,n} B_{m}^{n} \hat{O}_{m}^{n}(\bm{\hat{J}}_{i}) \label{eqCEF}
	\end{align}
	where $B_{m}^{n}$ denotes the CEF parameters and $\hat{O}_{m}^{n}$ symbolizes the Stevens operators, which are constructed based on the angular momentum $\bm{\hat{J}}$ after SOC.
	Accounting for the symmetry of the CEF environment, the number of terms in the Hamiltonian can be further reduced. 
	For triangular-lattice $ARECh_{2}$, within the $D_{3d}$ symmetry of $RECh_{6}$ octahedra, there are six remaining terms: $B_{2}^{0}$, $B_{4}^{0}$, $B_{4}^{3}$, $B_{6}^{0}$, $B_{6}^{3}$, and $B_{6}^{6}$. Excluding Ce$^{3+}$ and Sm$^{3+}$, there are three remaining terms: $B_{2}^{0}$, $B_{4}^{0}$, and $B_{4}^{3}$ \cite{stevens1952Matrix}.
	
	For any given compound, determining the CEF $B_{m}^{n}$ parameters is a crucial aspect.
	The most widely used method to estimate the $B_{m}^{n}$ parameters is based on the point charge model (PCM) \cite{hutchings1964PointCharge}.
	The PCM simplifies the complex environment around a magnetic ion by treating surrounding ligands as static point charges. This model calculates the electrostatic potential at the metal ion, which is used to determine the CEF parameters affecting the ion's electronic structure.
	Although the PCM lacks accuracy in quantitatively describing the CEF energy of rare-earth ions, it still serves as a useful rough approximation for semi-quantitative estimation of CEF energy levels.
	A more accurate approach is to directly calculate the excited states of the CEF using quantum chemical methods.
	Zangeneh \textit{et al.} \cite{PhysRevB.100.174436} performed quantum chemical calculations on Yb-based rare-earth chalcogenides, successfully reproducing the experimental observations.
	However, quantum chemical methods are time-consuming, and results can vary depending on the computational approach and basis sets used. Therefore, the more accurate determination of CEF energy levels for rare-earth compounds primarily relies on experimental methods.
	
	Inelastic neutron scattering (INS) and Raman scattering can both be used to probe CEF excitations in rare-earth compounds.
	Both are inelastic scattering techniques sharing formally similar scattering cross sections.
	INS is by far the most common method for studying CEF excitations in rare-earth materials. It can directly give the momentum-resolved CEF levels. While micron-sized samples can be detected with clear excitation signals by Raman scattering, INS requires a larger quantity of samples to obtain a better signal-to-noise ratio. Furthermore, Raman scattering offers higher energy resolution, allowing the study of phonons, other elementary excitations, and their potential couplings. This makes it possible to investigate CEF splitting under magnetic fields as well as the coupling between CEF excitations and phonons.
	
	Fig. \ref{fig:CEF} summarizes the CEF excitations reported for selected $ARECh_{2}$ compounds. Studies have predominantly focused on Ce-, Er-, Tm-, and Yb-based compounds, revealing a diverse landscape of CEF level structures. Clearly, the CEF energy levels are predominantly influenced by the $Ch$ site, with minimal impact from changes at the $A$ site. Meanwhile, the difference between R$\bar{3}$m and P6$_{3}$/mmc does not produce a significant difference, as indicated by comparison of the energy levels of CsErSe$_{2}$ or CsYbSe$_{2}$ with P6$_{3}$/mmc space group. These features are consistent with our understanding of the roles of the $A$ and $Ch$ sites in the crystal structure, despite slight adjustments.
	
	\textit{Ce-based compounds}.  
	According to Hund's rules, the ground-state spectral term for \ce{Ce^{3+}} with 4$f^{1}$ electron configuration is $^{2}F_{5/2}$ after SOC.
	Considering the $D_{3d}$ point group symmetry of the CEF, the $^{2}F_{5/2}$ spectral term further splits into three pairs of doubly degenerate Kramers states.
	However, all measured Ce-based compounds exhibited three CEF excitations, regardless of variations at the $A$-site or $Ch$-site, suggesting the presence of an extra mode beyond what was expected.
	Several extrinsic and intrinsic factors contributing to the extra mode have been discussed,
	for example, additional excitation peaks caused by impurities \cite{kulbakov2021Stripeyz,bordelon2021Magnetic,xie2023Stripe}, potential impacts of CEF-phonon coupling \cite{thalmeier1984Theory,bordelon2021Magnetic,adroja2012Vibron}, and transitions from 4$f$ electrons to the 5$s$ orbital \cite{loewenhaupt1979Dynamic}.
	However, none of them can give a valid explanation. The factors of local structural defects and the valence of Ce may be worth further investigation. 
	
	\textit{Er-based compounds}. 
	The 4$f$ orbital of \ce{Er^{3+}} accommodates 11 electrons, yielding a total spin quantum number $S$ of 3/2.
	Given that the orbital quantum number $L$ is 6 and the strong SOC characteristic of 4$f$ electrons, two spectral terms arise: $^{4}I_{15/2}$ with 16-fold degeneracy and $^{4}I_{13/2}$ with 14-fold degeneracy.
	The energy gap between the two spectral terms is around 0.8 eV \cite{Dieke:61}. Consequently, the spectral term $^{4}I_{15/2}$ denotes the SOC ground state configuration.
	From Fig. \ref{fig:CEF}, it is apparent that Er-based compounds are characterized by numerous and densely packed CEF excitations within the low-energy range.
	This characteristic ensures that for Er-based compounds, contributions from CEF excitations and van Vleck paramagnetism cannot be ignored even in lower measurement temperature ranges \cite{scheie2020Crystalfield,liu2024Finite}.
	
	\textit{Tm-based compounds}. 
	\ce{Tm^{3+}} with 4$f^{12}$ configuration has the spectral term $^{3}H_{6}$ featuring a 13-fold degeneracy and the spectral term $^{3}H_{4}$ with a 9-fold degeneracy after strong spin-orbit coupling (SOC). The remarkable energy difference of about 740 meV \cite{PETROV2019103} between the two terms suggests that one can safely focus on the lower $^{3}H_{6}$ spectral term when studying the magnetic properties below room temperature.
	Unlike the Kramers ions,  \ce{Tm^{3+}} with an even number of 4$f$ electrons has no Kramers doublets protected by time-reversal symmetry.
	The fundamental difference leads to the pronounced Ising-type behavior in Tm-based compounds \cite{PhysRevResearch.2.043013,PhysRevB.98.045119}.
	
	\textit{Yb-based compounds}. 
	CEF excitations in Yb-based compounds have been the most extensively studied among $ARECh_{2}$ family.
	The electron configuration of the 4$f$ orbital in \ce{Yb^{3+}} is 4$f^{13}$. The 4$f^{13}$ configuration emerges as a spectral term $^{2}I_{7/2}$ with an 8-fold degeneracy and another spectral term $^{2}I_{5/2}$ with a 10-fold degeneracy after SOC. The energy gap between the two spectral terms is about 1 eV \cite{PhysRevB.100.174436}, which means that we only need to consider the contribution associated with the spectral term $^{2}I_{7/2}$.
	\ce{NaYbSe2} \cite{zhang2021Crystalline} and \ce{NaYbS2} \cite{zhuo2024Magnetism} were among the first rare-earth chalcogenides to be investigated for CEF excitations using Raman scattering.
	These studies not only revealed CEF excitations in \ce{NaYbS2} and \ce{NaYbSe2} but also demonstrated coupling between the CEF excitations and phonons \cite{zhang2021Crystalline,zhuo2024Magnetism}.
	Pai \textit{et al.} further elucidated this phenomenon using a vibronic-bound-state model \cite{pai2022Mesoscale, pai2024Angular}.
	These observations pave the way for investigating CEF-phonon interactions, a phenomenon crucial for understanding the interplay between lattice dynamics and magnetic properties in rare-earth systems.
	
	The variety of CEF schemes observed across different rare-earth-based compounds within the $ARECh_{2}$ family underscores the richness of this material system for exploring quantum magnetism. Understanding these single-ion properties lays the foundation for investigating the collective magnetic behaviors and ground states in these materials, which will be explored in the following sections.
	
\section{Magnetic Hamiltonian and finite temperature magnetism}
\label{section:Finite}
	
	\begin{table*}[t]
		\caption{\textbf{Collected parameters related to the ground state of $\bm {ARECh_{2}}$.} The exchange parameters are based on the nearest-neighbor triangular-lattice anisotropic model (Eq. \ref{eqNNTL}) or the $J_{1}-J_{2}$ XXZ model, with values expressed in units of $\rm k_{\rm B}$. For comparison, the isotropic $J_{1}$ values are represented as $J_{\pm}=J_{1}/2$ and the anisotropy factor $\Delta=\frac{J_{zz}}{2J_{\pm}}$. The column labeled $T$ at Rln2 indicates the temperature at which the magnetic entropy $S_{m}$ approaches Rln2.}
		\label{table:Jpara}
		\begin{ruledtabular}
			\begin{tabular}{ccccccccccccc}
				Compounds & Ground state & $T_{\rm N}$ & $g_{ab}$ & $g_{c}$ &$J_{\pm}$&$J_{zz} $&$J_{\pm\pm}$&$J_{z\pm}$&$J_{2}/J_{1}$&$\Delta$& $T$ at Rln2 & Ref.\\
				\colrule
				
				KCeS$_{2}$ & stripe-$yz$ & 0.38 K & 1.65 & 0.6 &&&&&&&20 K& \cite{bastien2020Longrange, kulbakov2021Stripeyz} \\
				CsCeSe$_{2}$ & stripe-$yz$ & 0.35 K & 1.77 & (0.25) &0.42 K&0.21 K&0.44 K&0.34 K&&0.25&4 K& \cite{xie2024Quantum, xie2023Stripe} \\
				KErSe$_{2}$ & stripe-$x$ & 0.2 K &  &  & 0.12 K & 0.70 K & $-$0.46 K & 0.70 K &&2.92&2 K& \cite{xing2021Stripe, ding2023Stripe} \\
				KYbSe$_{2}$ & 120$^{\circ}$ & 0.29 K & 3.0 & 1.8 & 2.54 K &&&& 0.044 &&10 K& \cite{scheie2024Nonlinear, lee2024Magnetica} \\
				CsYbSe$_{2}$ & 120$^{\circ}$ & $<$\,0.4 K & 3.25 & 0.3 & 2.29 K & 4.53 K &&&0.029&0.99&15 K& \cite{xie2023Complete, xing2019Fieldinducedb} \\
				
				KYbO$_{2}$ & disordered & $<$\,70 mK & 3.08 & 3.08 &&&&&&&& \cite{grussler2023Roleb} \\
				NaYbO$_{2}$ & disordered & $<$\,70 mK & 3.294 & 1.726 & 2.96 K & 5.22 K &&&&0.88&35 K& \cite{bordelon2019Fieldtunablea, ding2019Gapless} \\
				NaYbS$_{2}$ & disordered & $<$\,50 mK & 3.14 & 0.86 &3.52 K&$-$0.87 K&$-$0.18 K&2.66 K&&$-$0.12&20 K& \cite{zhuo2024Magnetism, sarkar2019Quantum, sichelschmidt2019Electron} \\
				NaYbSe$_{2}$ & disordered & $<$\,20 mK & 3.10 & 0.96 & 3.27 K & $-$0.95 K &&&&$-$0.15& 15 K & \cite{zhang2021Effective, ranjith2019Anisotropic, zhang2022Lowenergya, dai2021Spinon, scheie2024Spectrum} \\
				&  &  &  & & 3.20 K &  &&& 0.071 && & \cite{scheie2024Nonlinear, scheie2024Spectrum} \\
				
				\colrule

				KTmSe$_{2}$ & disordered & $<$\,60 mK &&&& 3.37 K &\multicolumn{2}{c}{(TFIM, $h$ = 13.1 K)}&0.062&&15 K& \cite{zheng2023Exchangerenormalized} \\
				
			\end{tabular}
		\end{ruledtabular}
	\end{table*}

	In this section, we focus on the magnetism of $ARECh_{2}$ at finite temperatures.
	For the $ARECh_{2}$ crystals, the complete Hamiltonian can be expressed as \cite{zhang2021Effective}:
	\begin{align}
		{\mathcal H} =\ \mathcal H_{0} + \mathcal H_{\rm C} + \mathcal H_{\rm SOC} + \mathcal H_{\rm CEF} + \mathcal H_{\rm spin} \label{eqtotal}
	\end{align}
	where $\mathcal H_{0}$ is the sum of the kinetic and potential energies of the electrons, $\mathcal H_{\rm C}$ represents the Coulomb interaction term, $\mathcal H_{\rm SOC}$ is the SOC term, $\mathcal H_{\rm CEF}$ is the CEF contribution, and $\mathcal H_{\rm spin}$ is the exchange interaction between spins. It is observed that the SOC gaps for free $RE^{3+}$ ions are all above 200 K, with the exceptions of Sm and Eu \cite{Dieke:61}. Consequently, in most scenarios, it is sufficient to consider only the last two terms, which form an effective magnetic Hamiltonian describing thermodynamics at the temperatures of interest.
	 
	Having addressed the CEF Hamiltonian in Section \ref{section:Ion}, our focus now turns to the spin Hamiltonian $\mathcal H_{\rm spin}$.
	Regardless of which rare-earth-based compound is considered, the description of $\mathcal H_{\rm spin}$ based on total angular momentum $\bm{\hat{J}}$ is always applicable.
	For the triangular lattice, the spin Hamiltonian $\mathcal H_{\rm spin}$ for the angular momentum $\bm{\hat{J}}$ can be further simplified based on symmetry analysis. The interactions along the $\bm{\delta}_1$-bond allow four spatial symmetry operations: one is the $C_{2}$ rotational operation along the $\bm{\delta}_1$-bond, another is the $C_{3}$ rotational operation along the $c$-axis, and the translations $T_{1}$ and $T_{2}$ along the $\bm{\delta}_1$-bond and $\bm{\delta}_2$-bond directions, respectively.
	Considering these symmetry operations \cite{PhysRevB.94.035107,PhysRevX.9.021017}, the spin Hamiltonian $\mathcal H_{\rm spin}$ can be expressed as \cite{liu2024Finite}:
	\begin{align}
		& \mathcal H_{\rm spin} = \sum_{\langle i j\rangle} \left[ \vartheta_{z z} \hat{J}_i^z \hat{J}_j^z+ \vartheta_{ \pm}\left(\hat{J}_i^{+} \hat{J}_j^{-}+\hat{J}_i^{-} \hat{J}_j^{+}\right)\right. \notag \\
		& +\vartheta_{ \pm \pm}\left(\gamma_{i j} \hat{J}_i^{+} \hat{J}_j^{+}+\gamma_{i j}^* \hat{J}_i^{-} \hat{J}_j^{-}\right) \notag\\
		& \left.-\frac{i \vartheta_{z \pm}}{2}\left(\gamma_{i j} \hat{J}_i^{+} \hat{J}_j^z-\gamma_{i j}^* \hat{J}_i^{-} \hat{J}_j^z+\langle i \leftrightarrow j\rangle\right) \right]  \label{totalHam}
	\end{align}
	where $\hat{J}_{i}^{\alpha}$ ($\alpha = x, y, z$) is the components of angular momentum $\bm{\hat{J}}$ operators at site $i$ after SOC and $\hat{J}_{i}^{\pm} = \hat{J}_{i}^{x} \pm i\hat{J}_{i}^{y}$ are non-Hermitian ladder operators. The NN anisotropic spin interactions are denoted by $\vartheta_{zz}$, $\vartheta_{\pm}$, $\vartheta_{\pm\pm}$, and $\vartheta_{z \pm}$. The phase factor $\gamma_{i j}$ is taken as 1, $e^{i\frac{2\pi}{3}}$, $e^{-i\frac{2\pi}{3}}$ along the three bonds $\bm{\delta}_1,\bm{\delta}_2,$ and $\bm{\delta}_3$ [see Fig. \ref{fig:structure}(c)].
	
    For Kramers ions, when the thermal energy at the measurement temperature is significantly lower than the first excited CEF energy level, the ground state doublet remains well isolated.
	We can introduce an effective spin-1/2 $\bm{\hat{S}}_{i}$ that acts on the local ground state doublet. Therefore, the NN triangular-lattice anisotropic effective spin-1/2 Hamiltonian can be employed as follows \cite{PhysRevLett.115.167203,PhysRevB.94.035107}:
	\begin{align}
		{\mathcal H}_{\rm spin-1/2} =\ & \sum_{\langle ij\rangle}\left[J_{zz} \hat{S}_i^z \hat{S}_j^z+J_{\pm}\left(\hat{S}_i^{+} \hat{S}_j^{-}+\hat{S}_i^{-} \hat{S}_j^{+}\right)\right. \notag\\
		& +J_{\pm\pm}\left(\gamma_{ij} \hat{S}_i^{+} \hat{S}_j^{+}+\gamma_{ij}^* \hat{S}_i^{-} \hat{S}_j^{-}\right) \notag\\
		& \left.-\frac{iJ_{z\pm}}{2}\left(\gamma_{ij} \hat{S}_i^{+} \hat{S}_j^z-\gamma_{ij}^* \hat{S}_i^{-} \hat{S}_j^z+\langle i \leftrightarrow j\rangle\right)\right] \label{eqNNTL}
	\end{align}
	where $J_{zz}$ and $J_{\pm}$ are the bond-independent terms in the XXZ model, $J_{\pm\pm}$ and $J_{z\pm}$ are the bond-dependent anisotropic terms, $\hat{S}^{\pm}_i = \hat{S}^x_i \pm i \hat{S}^y_i$, and the phase factor $\gamma_{ij} = 1, e^{i 2\pi/3}, e^{-i2\pi/3}$ for the bond $ij$ along the $\bm{\delta}_1,\bm{\delta}_2,\bm{\delta}_3$ direction [see Fig. \ref{fig:structure}(c)].
	If the bond-dependent anisotropy is weak, i.e., $J_{\pm\pm}=J_{z\pm}=0$, then the effective spin-1/2 anisotropic Hamiltonian will reduce to the XXZ model. 
	Further if $J_{zz}=2J_{\pm}$, it will reduce to the isotropic Heisenberg model.
	In some $ARECh_{2}$ compounds, such as \ce{KYbSe2} \cite{scheie2023Proximate} and \ce{CsYbSe2} \cite{xie2023Complete}, the anisotropic spin interactions are weak, allowing the Heisenberg-like model, such as the $J_{1}-J_{2}$ Heisenberg model, to adequately describe the magnetic behavior at low temperatures.
	
	\begin{figure*}[t]
		\includegraphics[width=1\linewidth]{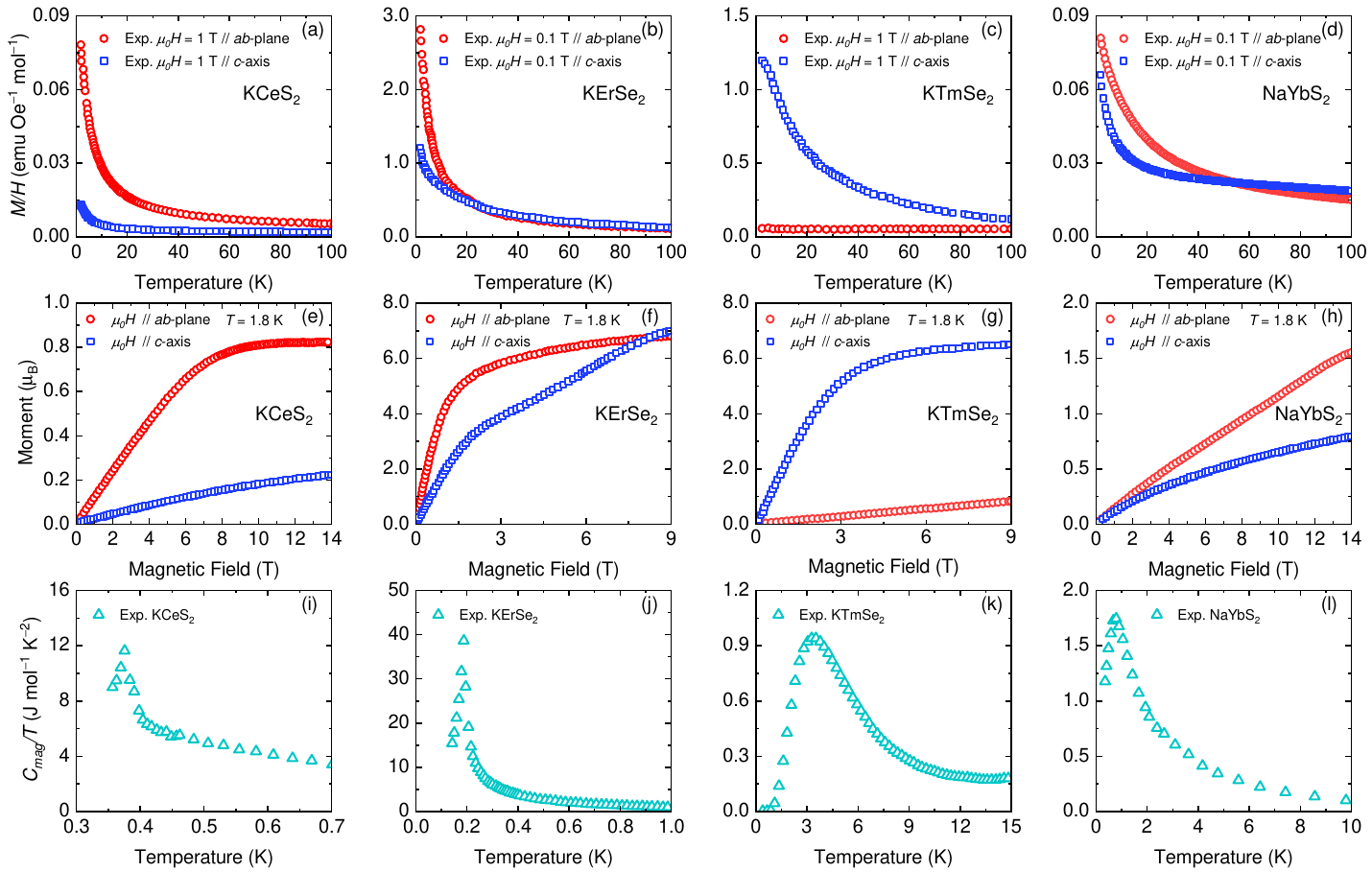}
		\caption{\label{fig:Mag} \textbf{Magnetization and magnetic specific heat of $\bm {ARECh_{2}}$.}
			(a)--(d) Temperature dependence of magnetization ($M/H$--$T$) for KCeS$_{2}$, KErSe$_{2}$, KTmSe$_{2}$, and NaYbS$_{2}$, respectively. (e)--(h) Field dependence of magnetization ($M$--$H$) at 1.8 K for the same compounds. (i)--(l) Temperature dependence of magnetic specific heat ($C_{mag}/T$) for each material. Data are taken from Ref. \cite{bastien2020Longrange, ding2023Stripe, zheng2023Exchangerenormalized, zhuo2024Magnetism}. 
		}
	\end{figure*}
	
	For non-Kramers ions, if we focus only on the ground state and first excited state of the CEF, we can construct a non-Kramers doublet NN spin Hamiltonian on the triangular lattice \cite{PhysRevB.98.045119}.
	\begin{align}
		{\mathcal H}_{\rm spin-1/2} =\ & \sum_{\langle ij\rangle}\left[J_{zz} \hat{S}_i^z \hat{S}_j^z+J_{\pm}\left(\hat{S}_i^{+} \hat{S}_j^{-}+\hat{S}_i^{-} \hat{S}_j^{+}\right)\right. \notag\\
		& +J_{\pm\pm}\left(\gamma_{ij} \hat{S}_i^{+} \hat{S}_j^{+}+\gamma_{ij}^* \hat{S}_i^{-} \hat{S}_j^{-}\right) \notag\\
		 \label{eq:non_kramers}
	\end{align}
	The spin Hamiltonian \ref{eq:non_kramers} is similar to Eq.\,\ref{eqNNTL}, but it is important to note that in Eq.\,\ref{eqNNTL}, all operators are dipolar, i.e., they conform to time-reversal antisymmetry. However, in the Hamiltonian \ref{eq:non_kramers}, only the $\hat{S}^{z}$ operator is dipolar, while $\hat{S}^{x}$ and $\hat{S}^{y}$ operators are multipolar and symmetric under time-reversal operation. Thus, the $\hat{S}^{z}$ operator does not directly couple with $\hat{S}^{x}$ (or $\hat{S}^{y}$) \cite{PhysRevResearch.2.043013}.
	Since traditional measurement techniques, such as magnetization and neutron diffraction, can only detect the contribution from dipolar magnetic moments, the resulting magnetic behavior exhibits strong Ising characteristics \cite{PhysRevResearch.2.043013,RN109}.
	
	If we consider the impact of an external magnetic field on magnetism, an additional Zeeman term ${\mathcal H}_{\rm Zeeman}$ must be included.
	\begin{align}
		{\mathcal H}_{\rm Zeeman} =\ -\mu_0 \mu_B \sum_i\left[g_{ab}\left(h_x \hat{S}_i^x+h_y \hat{S}_i^y\right)+g_c h_c \hat{S}_i^z\right] \label{eqZeeman}
	\end{align}
	where $g_{ab}$ and $g_{c}$ represent the $g$ factors in the $ab$-plane and along the $c$-axis, respectively; $h_{i}\ (i = x, y, z)$ denote the components of the external magnetic field.
	Through electron spin resonance (ESR), we can accurately determine the $g$ factors of the spin system \cite{sichelschmidt2019Electron,zhang2021Effective}.
	
	Based on the above discussion of the Hamiltonian, we will explore the magnetism at finite temperatures for several typical representatives within rare-earth chalcogenides.
	
	\textit{Ce-based compounds}. 
	Fig. \ref{fig:Mag}(a), (e) and (i) display the experimental data for the temperature-dependent magnetization ($M/H$--$T$) under a magnetic field of 1 T, magnetic field-dependent magnetization ($M$--$H$) at 1.8 K, and zero-field (ZF) magnetic specific heat over temperatures ($C_{mag}/T$) of \ce{KCeS2} \cite{bastien2020Longrange}, respectively.
	From the magnetization data, the magnetism of \ce{KCeS2} exhibits significant easy-plane characteristics, where the magnetic moments in the $ab$-plane are larger than those along the $c$-axis.
    This characteristic is also observed in \ce{CsCeSe2} \cite{xie2023Stripe}.
	Measurements of ZF $C_{mag}/T$ clearly reveal a $\lambda$-type phase transition peak near 0.35 K, indicating a transition from a paramagnetic state to an ordered state, which was further confirmed as a stripe-$yz$ ordered state \cite{kulbakov2021Stripeyz, xie2023Stripe}.
	Since the first excitation energy level of the CEF in Ce-based compounds is generally high, with even \ce{RbCeTe2} having nearly 30 meV \cite{ortiz2022Electronic}, an effective spin-1/2 model provides a good description of the magnetism at finite temperatures within the low temperature range.
	
	\textit{Er-based compounds}. 
	The experimental data for the $M/H$--$T$, $M$--$H$, and $C_{mag}/T$--$T$ of \ce{KErSe2} \cite{ding2023Stripe} are shown in Fig. \ref{fig:Mag}(b), (f) and (j), respectively.
	From Fig. \ref{fig:Mag}(b), it is evident that the $M/H$--$T$ curves for KErSe$_{2}$ in the $ab$-plane and along the $c$-axis are nearly identical above 10 K, which is primarily related to the weak CEF excitations in \ce{KErSe2}. At relatively high temperatures, the main contribution to magnetism comes from CEF excitations.
	Due to the large total angular momentum ($J = 15/2$) of \ce{Er^{3+}}, its magnetic moment is much larger than that of Ce-based and Yb-based compounds, which is clearly evident from the comparison of $M$--$H$ data.
	Er-based compounds also exhibit magnetic phase transitions, typically occurring at temperatures of just a few tenths of a Kelvin. ZF specific heat measurements can detect the transition from the paramagnetic state to the stripe-$x$ ordered state.
	Due to their weak CEF excitations, Er-based compounds are well-suited for spin Hamiltonians based on total angular momentum. For example, for \ce{KErTe2}, starting from a Hamiltonian based on total angular momentum and incorporating mean field (MF) theory can effectively simulate magnetization and specific heat data under different magnetic fields \cite{liu2024Finite}.
	
	\textit{Tm-based compounds}. 
	\ce{KTmSe2}, which has been studied in depth,\cite{zheng2023Exchangerenormalized} represents one of the few non-Kramers ions among rare-earth chalcogenides.
	From the $M/H$--$T$ [see Fig. \ref{fig:Mag}(c)] and $M$--$H$ [see Fig. \ref{fig:Mag}(g)] curves, it is evident that \ce{KTmSe2} exhibits strong Ising characteristics, with the magnetic moment along the $c$-axis being significantly larger than that in the $ab$-plane. This stands in sharp contrast to the magnetization of \ce{KCeS2}.
	The Ising magnetism in \ce{KTmSe2} is primarily determined by the symmetry of its CEF ground state and first excited state.
	By mapping, effective spin-1/2 operators derived from the CEF ground state and the first excited state enable the construction of a TFIM that captures the Ising magnetism.
	In \ce{KTmSe2}, a broad peak in $C_{mag}/T$--$T$ can be observed in the temperature range of 2--7 K, primarily attributed to the CEF first excited state.
	However, not all Tm-based compounds can be described by the TFIM. For example, \ce{NaTmTe2} requires consideration of spin exchange interactions within the $ab$-plane \cite{zheng2024Interplay}.
	
	\textit{Yb-based compounds}. 
	Yb-based rare-earth chalcogenides are currently the subsystem receiving the most attention \cite{schmidt2021Yb}.
	Extensive experimental data indicate that the ground states of \ce{NaYbO2}  \cite{bordelon2019Fieldtunablea}, \ce{NaYbS2} \cite{wu2022Magnetic}, and \ce{NaYbSe2}  \cite{dai2021Spinon,scheie2024Spectrum} are QSL states.
	The magnetization and magnetic specific heat data for these materials exhibit similar characteristics.
	Fig. \ref{fig:Mag}(d), (h), and (l) display the $M/H$--$T$, $M$--$H$, and ZF $C_{mag}/T$--$T$ data for \ce{NaYbS2} \cite{zhuo2024Magnetism}, respectively.
	Within the range of low temperatures and weak magnetic fields, the magnetization of \ce{NaYbS2} does not exhibit strong magnetic anisotropy compared to others. Therefore, the magnetism in some Yb-based compounds is also analyzed based on the $J_1-J_2$ Heisenberg model, as seen in \ce{KYbSe2} \cite{scheie2023Proximate,scheie2024Nonlinear} and \ce{CsYbSe2} \cite{xie2023Complete}.
	As the magnetic field increases, the $M$--$H$ curve of \ce{NaYbS2} exhibits strong anisotropy, and magnetization remains unsaturated up to 14 T, both along the $c$-axis and in the $ab$-plane.
	Magnetization measurements under high magnetic fields indicate that Yb-based compounds require a magnetic field of about 16 T for saturation in the $ab$-plane  \cite{zhang2021Effective}, while along the $c$-axis, saturation magnetization is not observed even up to 30 T \cite{zhang2021Effective}, suggesting strong van Vleck paramagnetism along the $c$-axis.
	No phase transition features are observed in the ZF specific heat of \ce{NaYbS2}  \cite{wu2022Magnetic,zhuo2024Magnetism}, except for a broad peak near 1 K, which is associated with spin exchange interactions.
	
	These magnetization and specific heat measurements at finite temperatures not only provide basic magnetic properties but also enable the determination of exchange interactions in the spin Hamiltonian and $g$ factors through data fitting.
	In Table \ref{table:Jpara}, we summarize the spin exchange parameters and some key characteristic temperatures reported in the literature, serving as a starting point for further discussion.
	 
	 \begin{figure*}[t]
	 	\includegraphics[width=1\linewidth]{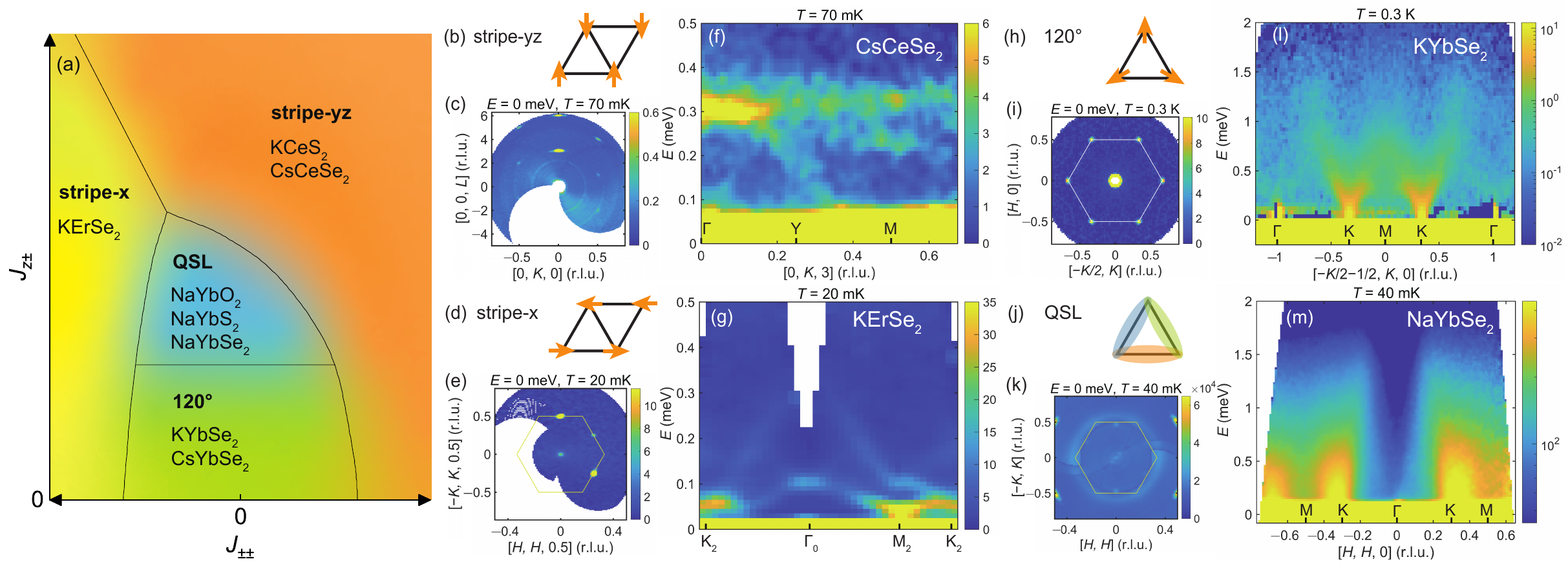}
	 	\caption{\label{fig:INS}  \textbf{Ground states and excitations in  $\bm {ARECh_{2}}$.} (a) A schematic phase diagram of the NN TL anisotropic model \cite{PhysRevB.94.035107, zhu2018Topography}. 
	 	(b) The arrangement of magnetic moments in the stripe-$yz$ order on a triangular lattice.  [(c), (f)] The magnetic Bragg peaks from elastic neutron scattering (ENS) and the low-energy spin excitations from inelastic neutron scattering (INS) of \ce{CsCeSe2} at 70 mK, with a ground state of stripe-$yz$ order, respectively. The peaks at (0, 0, 3) and (0, 0, 6) are nuclear Bragg peaks.
	 	(d) The arrangement of magnetic moments in the stripe-$x$ order on a triangular lattice. [(e), (g)] The magnetic Bragg peaks from ENS and the low-energy spin excitations  from INS of \ce{KErSe2} \cite{ding2023Stripe} at 20 mK, with a ground state of stripe-$x$ order, respectively. Spin wave excitations near $M$ point caused by magnetic ordering can be clearly observed.
	 	(h) The arrangement of magnetic moments in the AFM-120° order on a triangular lattice. [(i), (l)] The magnetic Bragg peaks from ENS and the low-energy spin excitations from INS of \ce{KYbSe2} \cite{scheie2023Proximate} at 0.3 K, with a ground state of AFM-120° order, respectively.
	 	(j) Schematic diagram of the short-range resonating valence bonds formed in the QLS phase. [(k), (m)] The ENS and the low-energy spin excitations from INS of \ce{NaYbSe2} \cite{dai2021Spinon} at 40 mK, with a QSL ground state, respectively. The characteristics of continuous excitations can be clearly observed throughout the Brillouin zone.}
	 \end{figure*}

\section{Ground state and low-energy spin excitations}
	\label{section:GS}
	
	In rare-earth chalcogenides $ARECh_{2}$, the ground state magnetism and related low-energy spin excitations are of particular interest.
	Neutron scattering is one of the most advantageous experimental techniques for studying ground state magnetism and low-energy spin excitations.
	Currently, research on $ARECh_{2}$ primarily focuses on rare-earth ions with an odd number of electrons.
	The ground state magnetism and low-energy spin excitations of these $ARECh_{2}$ compounds can be well described by the Hamiltonian presented in Eq. \ref{eqNNTL}.
	Therefore, in this section, we summarize the ground state magnetism and low-energy excitations of $ARECh_{2}$ compounds with an odd number of electrons.
	
	For the Hamiltonian described by Eq. \ref{eqNNTL}, there are existing theoretical studies including methods such as density matrix renormalization group (DMRG) \cite{PhysRevLett.119.157201,PhysRevX.9.021017,PhysRevB.95.165110,zhu2018Topography}, Monte Carlo simulations  \cite{PhysRevB.94.035107}, exact diagonalization (ED) \cite{PhysRevB.95.165110,zhu2018Topography,PhysRevB.103.205122}, and spin wave theory \cite{PhysRevB.94.035107}.
	The ground-state phase diagram, summarized based on theoretical calculations, is shown in Fig. \ref{fig:INS}(a).
	The phase diagram is clearly divided into four regions: stripe-$yz$, stripe-$x$, 120°, and QSL.
	The illustrations for these four phases can be found in Fig. \ref{fig:INS}. For the ordered phases, the magnetic moment configurations on the triangular lattice are explicitly depicted. The QSL phase, characterized by the absence of long-range magnetic order, is represented conceptually.
	In Ce-, Er-, and Yb-based compounds, all four corresponding phases can be found.
	This also fully demonstrates that Eq. \ref{eqNNTL} is not just a formal theory, but a robust framework that closely aligns with real materials, providing a predictive model that captures their complex magnetic behaviors and low-energy spin excitations accurately.
	
	\textit{Stripe-yz phase}. 
	The majority of Ce-based compounds, such as \ce{KCeO2} \cite{bordelon2021Magnetic}, \ce{KCeS2} \cite{bastien2020Longrange,avdoshenko2022Spinwave,kulbakov2021Stripeyz}, and \ce{CsCeSe2} \cite{xie2024Quantum,xie2023Stripe}, develop a long-range magnetic order.
	However, \ce{RbCeO2} is an exception, with no magnetic ordering observed even down to 60 mK.
	Neutron diffraction confirmed that the ground state of \ce{KCeS2} \cite{kulbakov2021Stripeyz} and \ce{CsCeSe2} \cite{xie2023Stripe} will form a stripe-$yz$ order with propagation vector $\textbf{k}=$ ($0$, $\bar{\frac{1}{2}}$, $\frac{1}{2}$) and ($0$, $\frac{1}{2}$, $1$), respectively.  
	Avdoshenko \textit{et al.} for the first time, performed INS on polycrystalline samples of \ce{KCeS2} \cite{avdoshenko2022Spinwave}, revealing low-energy spin excitations associated with the stripe-$yz$ phase, characterized by a sharp spin-wave mode centered at 0.34 meV for low $|Q|$, with an intensity maximum near $Q = 0$.
	Xie \textit{et al.} conducted a more detailed neutron scattering study on single crystal samples of \ce{CsCeSe2} \cite{xie2024Quantum,xie2023Stripe}.
	Fig. \ref{fig:INS}(f) illustrates the zero-field INS spectrum, which exhibits three key features:  a sharp excitation at the $\Gamma$ point, a nearly gapless acoustic mode, and two weakly dispersive optical modes. Comprehensive analysis of this spectrum revealed strong signatures of bond-dependent Kitaev interactions, providing crucial experimental evidence for theoretically predicted Kitaev-like physics in rare-earth-based materials.
	
	\textit{Stripe-x phase}. 
	In Er-based compounds, the ground state typically exhibits stripe-$x$ order.
	Due to the influence of CEF excitations over a wide temperature range and the low magnetic transition temperatures in Er-based compounds, there is limited research on their ground state magnetism.
	Ding \textit{et al.} were the first to investigate the ground state magnetism and low-energy spin excitations in \ce{KErSe2} using neutron scattering techniques \cite{ding2023Stripe}.
	Elastic neutron scattering (ENS) studies reveal long-range order below $T_{N} =$ 0.2 K with a propagation vector $\textbf{k}=$ ($\frac{1}{2}$, 0, $\frac{1}{2}$), as shown in Fig. \ref{fig:INS} \cite{ding2023Stripe}. 
	INS measurements show well-defined spin-wave excitations with a $\sim$0.03 meV gap at $M$ points in the Brillouin zone \cite{ding2023Stripe}. 
	Both the stripe-$x$ order and spin-wave excitations could be quantitatively understood from the anisotropic spin model (see Eq. \ref{eqNNTL}).
	
	\textit{120$^{\circ}$ phase}. 
	The 120° ordered state represents a unique magnetic structure.
	Despite forming long-range magnetic order, it still exhibits strong quantum fluctuations and even confined spinon excitations \cite{zhuo2024Magnetism}.
	\ce{KYbSe2} is a typical example of a compound that forms a 120° ordered state \cite{scheie2023Proximate} with the propagation vector $\textbf{k}=$ ($\bar{\frac{1}{3}}$, $\frac{1}{3}$, 0) \cite{scheie2024Nonlinear}. 
	More importantly, INS reveals a diffuse continuum with a sharp lower bound suggestive of fractionalized spinons (Fig. \ref{fig:INS}). With $J_{2}/J_{1} \sim 0.047$, KYbSe$_{2}$ sits remarkably close to the predicted quantum critical point (QCP) at $J_{2}/J_{1} \sim 0.06$  \cite{scheie2024Nonlinear,scheie2023Proximate}. 
	Xie \textit{et al.} conducted neutron scattering studies on another Yb-based compound, \ce{CsYbSe2}, which also forms an 120° ordered state \cite{xie2023Complete}. 
	Compared to \ce{KYbSe2}, CsYbSe$_{2}$ exhibits a more robust 120$^{\circ}$ order below 0.4 K, attributed to its smaller $J_{2}/J_{1} \sim 0.03$ \cite{xie2023Complete}. 
	These materials provide valuable insights into the emergence of magnetic order from highly fluctuating quantum states in frustrated magnets.
	
	\textit{QSL phase}. 
	The search for QSL states in $ARECh_{2}$ compounds represents a frontier in quantum magnetism, with potential implications for the development of novel quantum materials. \ce{NaYbSe2} has emerged as the most extensively studied QSL candidate within this family, serving as a prototype for investigating the hallmarks of quantum spin liquids in real materials.
	Initial studies of \ce{NaYbSe2} using magnetization and specific heat measurements revealed no magnetic phase transitions. \cite{ranjith2019Anisotropic,zhang2021Effective}.
	Dai \textit{et al.} first performed neutron scattering on single crystals, with ENS showing no magnetic Bragg peaks down to 40 mK, as shown in \ref{fig:INS}(k) \cite{dai2021Spinon}.
	More significantly, INS revealed a V-shaped excitation near the $\Gamma$ point, potentially indicating Fermi surface-like spinon excitations, see Fig. \ref{fig:INS}(m) \cite{dai2021Spinon}.
	Recently, Scheie \textit{et al.} combined INS and ultra-low temperature AC susceptibility measurements down to 20 mK, uncovering a 2.1 $\mu$eV energy gap in \ce{NaYbSe2}, suggesting a gapped $Z_{2}$ spin liquid ground state \cite{scheie2024Spectrum}.
	Bordelon \textit{et al.} and Wu \textit{et al.} have conducted INS studies on polycrystalline samples of \ce{NaYbO2} \cite{bordelon2019Fieldtunablea} and \ce{NaYbS2} \cite{wu2022Magnetic}, respectively.
	These results also indicate that the ground states of \ce{NaYbO2} and \ce{NaYbS2} are spin liquid states, though their precise nature remains to be determined.
	
\begin{figure}[t]
		\includegraphics[width=1\linewidth]{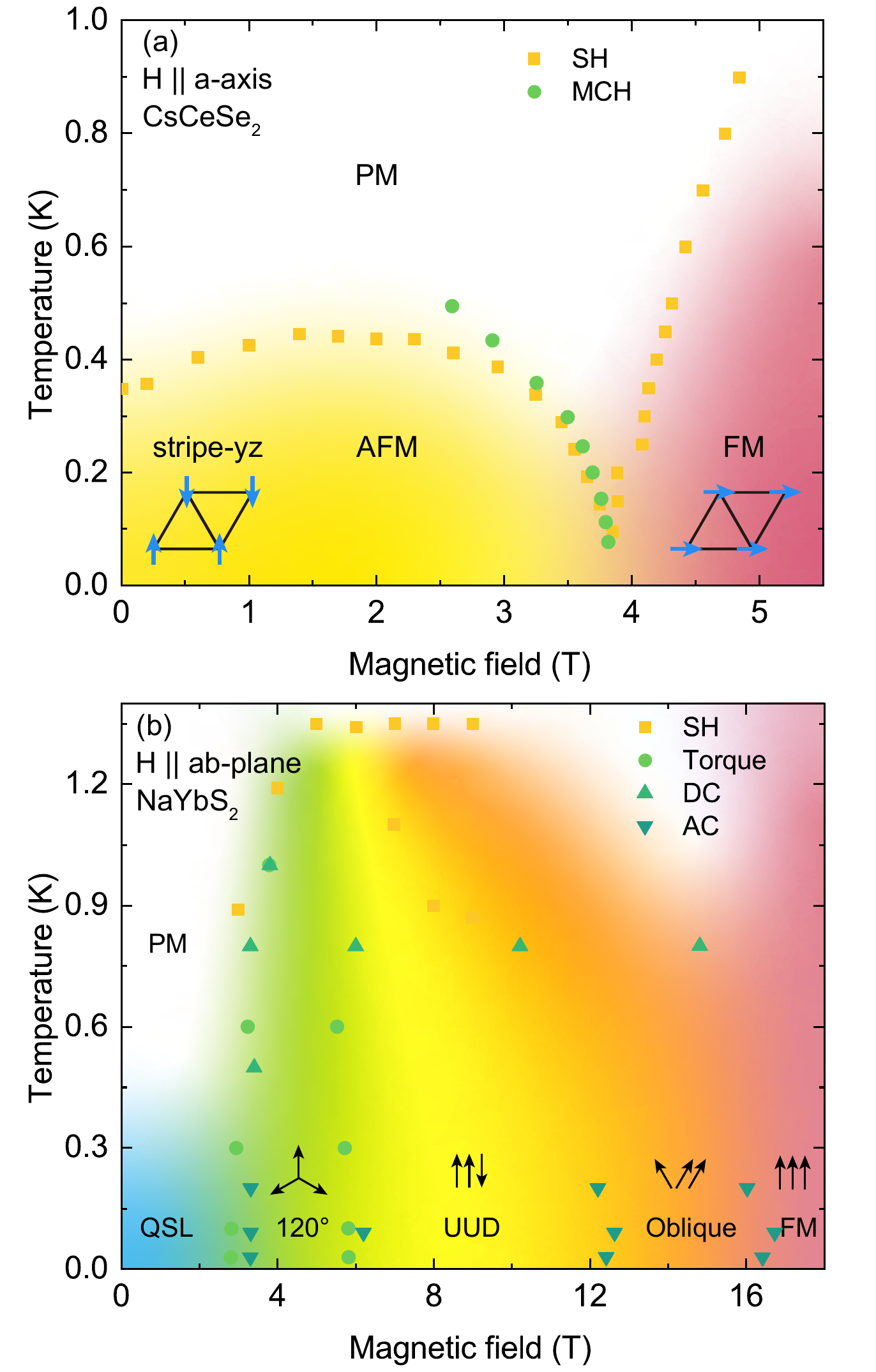}
		\caption{\label{fig:Phase}  \textbf{Magnetic phase diagrams of $\bm {ARECh_{2}}$.} (a) Temperature versus magnetic field phase diagram for CsCeSe$_{2}$ with field applied along the $a$-axis,determined from specific heat (SH) and magnetocaloric effect (MCE) measurements, showing paramagnetic (PM), stripe-$yz$, antiferromagnetic(AFM) and ferromagnetic (FM) phases \cite{xie2023Stripe}. (b) Phase diagram for NaYbS$_{2}$ with field in the $ab$-plane, determined from SH, torque measurements, and DC/AC susceptibility. Phases include QSL, 120° order, up-up-down (UUD), oblique, and field-polarized FM states \cite{wu2022Magnetic}.
		}
	\end{figure}

	Recently, there have been significant advancements in the study of quantum phase transitions in $ARECh_{2}$ under magnetic fields, primarily focusing on Ce-based and Yb-based compounds.
	Fig. \ref{fig:Phase}(a) presents the magnetic field-temperature ($H$--$T$) phase diagram of \ce{CsCeSe2} with the magnetic fields applied along the $a$-axis \cite{xie2023Stripe}.
	The QCP from the stripe-$yz$ to the fully polarized state occurs at a magnetic field of 3.86 T.
	Compared to Ce-based compounds, Yb-based compounds exhibit a richer variety of magnetic field-induced phase transitions.
	As shown in Fig. \ref{fig:Phase}(b), we use \ce{NaYbS2} as an example to illustrate a series of magnetic structures induced by a magnetic field in the $ab$-plane \cite{wu2022Magnetic}.
	As the magnetic field increases, the ground state of \ce{NaYbS2} sequentially transits through the QSL, 120°, up-up-down (UUD), oblique, and spin-polarized states.
	Field-induced phase transitions in \ce{NaYbSe2} \cite{zhang2021Effective,ranjith2019Anisotropic}, with the magnetic fields in the $ab$-plane, also exhibit characteristics similar to those observed in \ce{NaYbS2}.
	However, the quantum critical behavior at the phase boundaries remains largely unexplored. These quantum critical regions may exhibit novel topological properties, deserving further investigation.
	
\section{Outlook}
\label{section:Outlook}

	Research into the $ARECh_{2}$ family has advanced rapidly since 2018, opening new avenues for exploring frustrated magnetism. This field continues to develop swiftly, presenting both opportunities and challenges. Following the narrative of this review, we can naturally see several interesting and noteworthy directions:
	
	\textit{QCPs}. In-plane magnetic fields can induce novel quantum phases from different ground states. The QCPs within these phase transitions often exhibit enhanced quantum fluctuations and unusual scaling behavior. Probing the excitation spectra and thermodynamic properties near these critical points could reveal fundamental aspects of quantum criticality in frustrated magnets.
	
	\textit{Topological Properties}. QCPs and QSL phases often host novel topological properties and excitations. Thermal Hall effect measurements have emerged as powerful probes of these features, offering insights into the fractional nature and topological character \cite{savary2017Quantum, zhang2024Thermal, cairns2024Gapped}. However, the complexity of QSLs demands further development of this and complementary techniques. Advanced thermal transport measurements, combined with other probes, could provide a more complete picture of exotic quasiparticles and emergent gauge fields in QSLs.
	
	\textit{Entanglement Witnesses}. As a key feature of QSLs, the highly entangled ground state requires further development of experimental probes. The most promising approach to extracting information about entanglement in materials is through entanglement witnesses \cite{scheie2023Proximate, laurell2024Witnessing, sabharwal2024Witnessing}. This advanced method is under active development and merits further refinement and applications to real materials and experiments.
	
	\textit{Material Tunability}. The high tunability of the $ARECh_{2}$ family merits further exploration of the evolution between ordered and disordered phases. $A$-site modulation in Yb-based materials can induce evolution from 120° order to QSL states. Similar evolution may potentially be realized in Ce- and Er-based compounds.
	Additionally, dilute magnetic doping of non-magnetic analogs enables tracking the evolution of spin correlations from isolated paired magnetic ions to many-body entangled states \cite{haussler2022Diluting, pritchardcairns2022Tracking, cairns2024Gapped}. This approach bridges single-ion and collective magnetic behavior.
	Systematic investigation of such element-tuned evolution, combined with anisotropic Hamiltonian descriptions, holds significant potential for uncovering new insights into the emergent phenomena of quantum magnetism.
	Moreover, research on Ce-, Er-, Tm-, and Yb-based compounds has provided valuable paradigms and insights into frustrated magnetism in triangular-lattice rare-earth magnets. Expanding studies to other rare-earth-based members in the $ARECh_{2}$ family presents exciting opportunities to uncover additional exotic magnetic phases.
	
	\textit{Superconductivity}. Despite lacking a smoking-gun QSL signature, direct applications of QSL candidates remain attractive. \ce{NaYbSe2} pioneered this approach, exhibiting superconductivity at $\sim$\,5.6 K under pressure \cite{GuoLin2023, jia2020Mott, zhang2020Pressure}. Further investigation of Te-based compounds, with potentially smaller gaps amenable to doping via techniques like ionic liquid gating, could be promising for achieving superconductivity.
	
	These directions highlight the rich potential of the rare-earth chalcogenides $ARECh_{2}$ family for fundamental studies, pointing towards the frontiers of research in frustrated magnetism.

\section{Acknowledgements}
This work was supported by the National Key Research and Development Program of China (Grant No. 2022YFA1402704), 
the National Science Foundation of China (Grant No. 12274186), the Strategic Priority Research Program of the Chinese Academy of Sciences (Grant No. XDB33010100), and the Synergetic Extreme Condition User Facility (SECUF).
The authors thank Jun Zhao (Fudan University), Xingye Lu (Beijing Normal University), and Tao Xie (Sun Yat-sen University) for sharing valuable neutron data.

%

\end{document}